\documentclass[12pt]{article}

\usepackage[numbers]{natbib}
\usepackage{graphicx}
\usepackage{amsthm}
\usepackage{amsmath}
\usepackage{amssymb}

\newcommand\pubnumber{NuPhys2015-Graf}
\newcommand\pubdate{\today}

\def\napoli{University College London, Gower Street, London WC1E 6BT, UK}
\def\support{\footnote{lukas.graf.14@ucl.ac.uk}}

\def\Title#1{\begin{center} {\Large #1 } \end{center}}
\def\Author#1{\begin{center}{ \sc #1} \end{center}}
\def\Address#1{\begin{center}{ \it #1} \end{center}}

\newcommand\pubblock{\rightline{\begin{tabular}{l} \pubnumber\\
         \pubdate  \end{tabular}}}
\newenvironment{Abstract}{\begin{quotation}  }{\end{quotation}}
\newenvironment{Presented}{\begin{quotation} \begin{center} 
             PRESENTED AT\end{center}\bigskip 
      \begin{center}\begin{large}}{\end{large}\end{center} \end{quotation}}
\def\Acknowledgements{\bigskip  \bigskip \begin{center} \begin{large}
             \bf ACKNOWLEDGEMENTS \end{large}\end{center}}

\def\beq{\begin{equation}}
\def\eeq#1{\label{#1}\end{equation}}
\def\eeqn{\end{equation}}

\def\beqa{\begin{eqnarray}}
\def\eeqa#1{\label{#1}\end{eqnarray}}
\def\eeqan{\end{eqnarray}}

\let\bar=\overbar

\def\Dslash{\not{\hbox{\kern-4pt $D$}}}
\def\dslash{\not{\hbox{\kern-2pt $\del$}}}

\def\msb{{\bar{\ssstyle M \kern -1pt S}}}

\begin{document}
\begin{titlepage}
\pubblock

\vfill
\Title{Falsifying Baryogenesis with Neutrinoless Double Beta Decay}
\vfill
\Author{ Lukas Graf \support}
\Address{\napoli}
\vfill
\begin{Abstract}
We discuss the relation between lepton number violation at high and low energies, particularly, the constraints on baryogenesis models, which would be implied by an observation of neutrinoless double beta decay. The primordial baryon asymmetry can be washed out by effective lepton number violating operators triggering neutrinoless double beta decay in combination with sphaleron processes. A generic conclusion is that popular models of baryogenesis are excluded if a non-standard mechanism of neutrinoless double beta decay, i.e., other than the standard light neutrino exchange, is observed. Apart from the effective field approach, we also outline the possible extension of our arguments to a general UV-completed model.

\end{Abstract}
\vfill
\begin{Presented}
NuPhys2015, Prospects in Neutrino Physics\\
Barbican Centre, London, UK,  December 16--18, 2015
\end{Presented}
\vfill
\end{titlepage}
\def\thefootnote{\fnsymbol{footnote}}
\setcounter{footnote}{0}

\section{Introduction}
At present, the theoretically most acceptable solution to the neutrino mass problem is the concept of Majorana neutrinos. 
In case that this theoretical construction is really the mechanism how neutrinos obtain their masses in nature, we expect to see lepton number violating (LNV) processes, particularly neutrinoless double beta ($0\nu\beta\beta$) decay \cite{Furry}. Although it is not possible to pinpoint a model just on the basis of observation of this single process, a number of interesting conclusions can be drawn. In this text we focus on cosmological implications of effective LNV operators that occur in general mechanisms of Majorana neutrino mass generation \cite{Deppisch}. 

\section{Effective Approach to $0\nu\beta\beta$}
\begin{figure*}[t!]
\centering
\begin{minipage}[t]{0.45\linewidth}
\centering
\includegraphics[scale=0.315]{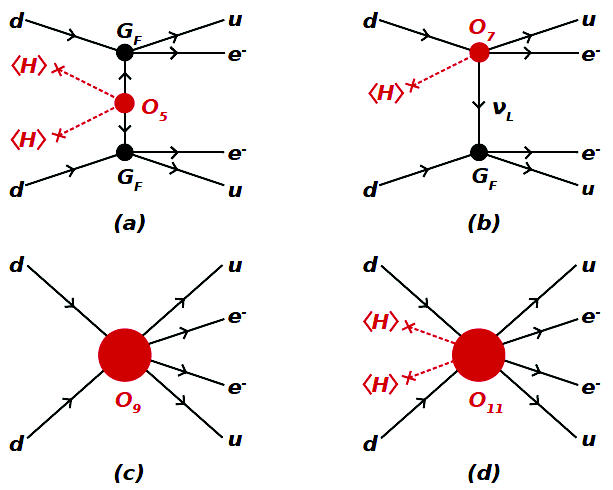}
\caption{Diagrams showing the contributions of the operators \eqref{eq:ops} to $0\nu\beta\beta$ decay.}
\label{fig:1}
\end{minipage}
\quad
\begin{minipage}[t]{0.46\linewidth}
\centering
\includegraphics[scale=0.18]{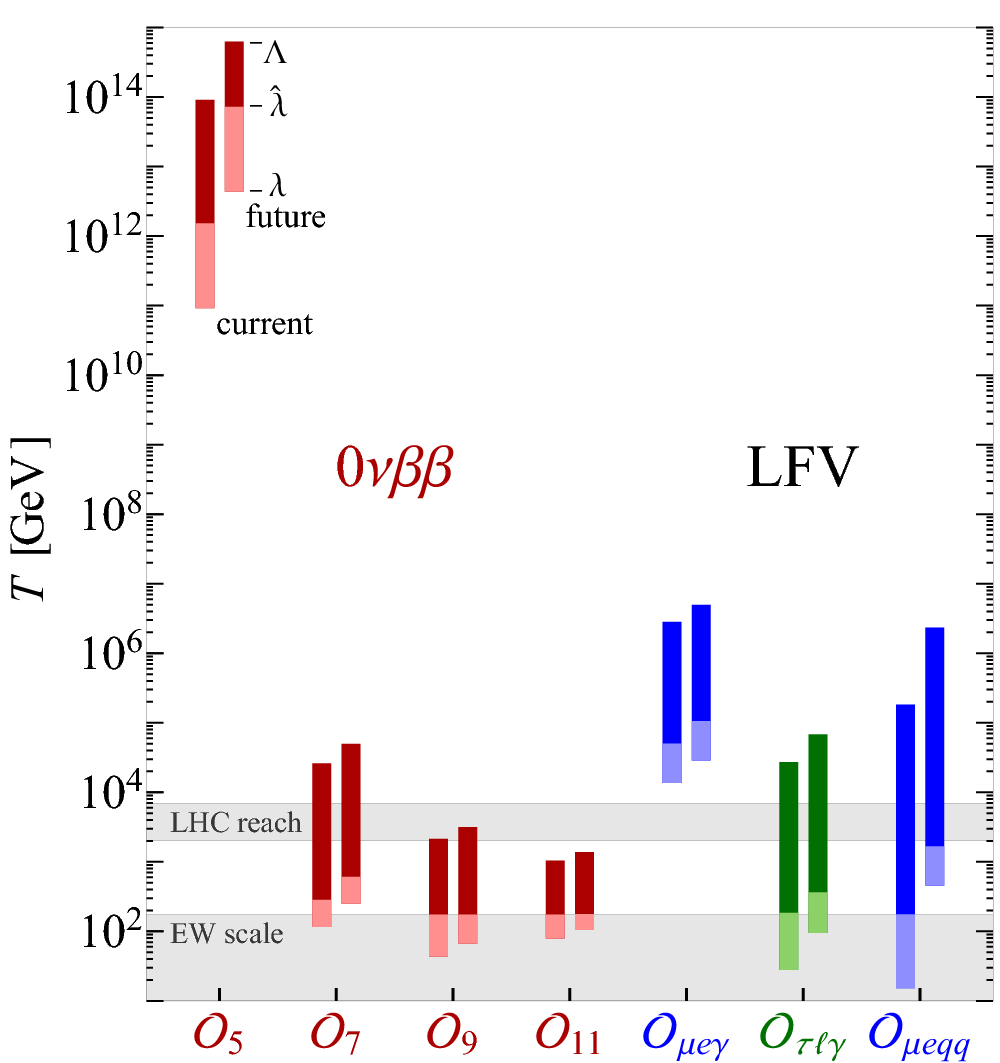}
\caption{Temperature intervals in which the LNV and LFV operators are in equilibrium assuming observation at the current/future (left/right bars) sensitivity.}
\label{fig:2}
\end{minipage}
\end{figure*}

If we assume that all the new physics lives high above the electroweak scale, then regardless of the details of the new high-scale theory all the new phenomena occurring below electroweak scale are described by higher-dimensional effective operators. Hence, observable low scale manifestations of all high-energy models leading to small Majorana neutrino masses are just consequences of effective operators which break the $B-L$ number. The LNV processes which are most relevant for $0\nu\beta\beta$ violate lepton number by two units and conserve baryon number. Hence, we can focus on odd-dimensional $\Delta L=2$ effective operators. Up to dimension 11 there is a list of 129 LNV operators \cite{deGouvea}, from which we will concentrate on the following four examples
\begin{eqnarray}
\mathcal{O}_5 = \frac{1}{\Lambda_5}(L^iL^j)H^k H^l \varepsilon_{ik} \varepsilon_{jl},  &\ &
\mathcal{O}_7 = \frac{1}{\Lambda_7^3}(L^id^c)(\bar{e^c}\bar{u^c})H^j \varepsilon_{ij}, \nonumber \\
\mathcal{O}_9 = \frac{1}{\Lambda_9^5}(L^iL^j)(\bar{Q_i}\bar{u^c})(\bar{Q_j}\bar{u^c}),  &\ &
\mathcal{O}_{11} = \frac{1}{\Lambda_{11}^7} (L^iL^j)(Q_k d^c)(Q_l d^c)H_m\bar{H}_i \varepsilon_{jk} \varepsilon_{lm}. \label{eq:ops}
\end{eqnarray}
Here, $L=\left(\nu_L, e_L\right)^T$ is the $SU(2)_L$ lepton doublet, $Q=\left(u_L, d_L\right)^T$ is the $SU(2)_L$ quark doublet and $H=(H^+, H^0 )^T$ is the $SU(2)_L$ Higgs doublet, whose neutral component acquires a non-zero vacuum expectation value breaking the electroweak gauge symmetry. The fields $\bar{e^c}, \bar{u^c}$ and $\bar{d^c}$ are the charge conjugates of the $SU(2)_L$ singlet right-handed charged-fermion operators. The contributions of these operators to $0\nu\beta\beta$ decay are shown in Fig. \ref{fig:1}.

The theoretical formula for the $0\nu\beta\beta$ decay half life $T_{1/2}$ reads
\begin{equation} \label{eq:halflife}
T^{-1}_{1/2}=\epsilon_i^2G_i|M_i|^2,
\end{equation}
where $G_i$ is the decay phase space factor and $M_i$ stands for the matrix element for a given isotope and operator, which is assumed to be dominant. The coefficient $\epsilon_i$ is an effective coupling of a specific operator. The current bounds on the $0\nu\beta\beta$ decay half life given by experimental searches in $^{76}$Ge and $^{136}$Xe are $T_{1/2}>2.1\times10^{25}$ y \cite{Agostini} and $T_{1/2}>1.9\times10^{25}$ y \cite{Gando} (at $90\%$ C. L.), respectively.  The planned future sensitivity could be improved by two orders of magnitude to $T_{1/2}\approx 10^{27}$ y \cite{Gomez}.

Relating the scales of the four operators in equation \eqref{eq:ops} to the corresponding effective couplings we get
\begin{equation} \label{eq:opscales}
m_e\epsilon_{5}=\frac{g^2v^2}{\Lambda_5}, \ \ \ \frac{G_F\epsilon_{7}}{\sqrt{2}}=\frac{g^3v}{2\Lambda_7^3},\ \ \
\frac{G_F^2\epsilon_{\{9,11\}}}{2m_p}=\Bigg\{\frac{g^4}{\Lambda_9^5},\frac{g^6v^2}{\Lambda_{11}^7}\Bigg\},
\end{equation}
where $G_F$ is the Fermi coupling, $m_e$ and $m_p$ denotes the electron and proton mass, respectively, and $v$ represents the Higgs vacuum expectation value. Moreover, a generic (average) coupling constant $g$ was included just to show the estimated scaling in a tree level UV completion of the operator.

If we employ the relations \eqref{eq:halflife} and \eqref{eq:opscales}, we can calculate the current upper bounds on the scales of the considered effective operators $\Lambda_D$. The numerical values are depicted in Fig. \ref{fig:2}.

\section{Falsification of High-scale Baryogenesis?}
The operators in \eqref{eq:ops} will not only induce $0\nu\beta\beta$ decay, but in connection with sphaleron processes \cite{Klinkhamer} (translating the asymmetry from leptons to baryons) they also lead to the washout of a $B-L$ asymmetry created in a baryogenesis mechanism. The Boltzmann equation describing the net lepton number evolution in dependence on temperature $T$ for a single LNV  $\Delta L=2$ operator of dimension $D$ reads \cite{Deppisch}
\begin{equation} \label{eq:boleq}
n_{\gamma}HT\frac{\mathrm{d}\eta_L}{\mathrm{d}T}=c_D\frac{T^{2D-4}}{\Lambda_D^{2D-8}}\eta_L.
\end{equation}
Here $n_{\gamma}\approx2T^3/\pi^2$ denotes the equilibrium photon density and $H\approx1.66\sqrt{g_*}T^2/\Lambda_{Pl}$ represents the Hubble parameter with the effective number of degrees of freedom $g_*$ and the Planck scale $\Lambda_{Pl}=1.2\times10^{19}$ GeV. The constant $c_D$ depends on the considered operator. The condition for a process to be in equilibrium is
\begin{equation}
\frac{\Gamma_W}{H}\equiv\frac{c_D}{n_{\gamma}H}\frac{T^{2D-4}}{\Lambda_D^{2D-8}}\approx 0.3c_D\frac{\Lambda_{Pl}}{\Lambda_D}\left(\frac{T}{\Lambda_D}\right)^{2D-9} \gtrsim 1.
\end{equation}
This inequality simply requires the decay rate to be large in comparison with expansion rate of the universe, otherwise the considered process would depart from equilibrium. This is here satisfied whenever the temperature $T$ lies in the interval
\begin{equation}
\Lambda_D \gtrsim T \gtrsim \lambda_D\equiv \Lambda_D\left(\frac{\Lambda_D}{0.3c_D\Lambda_{Pl}}\right)^{\frac{1}{2D-9}}.
\end{equation}
The lower limit $\lambda_D$ therefore represents the temperature above which any pre-existing lepton number asymmetry will be washed out in case that $0\nu\beta\beta$ decay is observed at the corresponding rate and if the given operator $\mathcal{O}_D$ gives the dominant contribution. On the other hand, the scale $\Lambda_D$ is the upper limit given by the validity of the effective operator approach, as above this scale the UV-completed model must be considered.

The precise lower limit for the washout scale $\hat{\lambda}_D$, above which the typical primordial asymmetry of order one can be suppressed down to the electroweak scale, can be determined by solving the Boltzmann equation \eqref{eq:boleq}.

Figure \ref{fig:2} shows that if a non-standard $0\nu\beta\beta$ decay is observed, then high-scale baryogenesis can be excluded. However, in case that $0\nu\beta\beta$ decay is dominated by the standard mass mechanism, the origin of neutrino masses and baryogenesis are most probably high-scale phenomena. The crucial point is distinguishing among various mechanisms, which can be attained e.g. through the observation of neutrinoless double beta decay in multiple isotopes or the measurement of the decay distribution. Although $0\nu\beta\beta$ decay can probe LNV only in the first lepton generation, if lepton flavour violation is observed, our argumentation can be extended to the washout of other flavours - see Fig. \ref{fig:2}. 

\begin{figure}
  \begin{minipage}[c]{0.6\textwidth}
    \includegraphics[width=\textwidth]{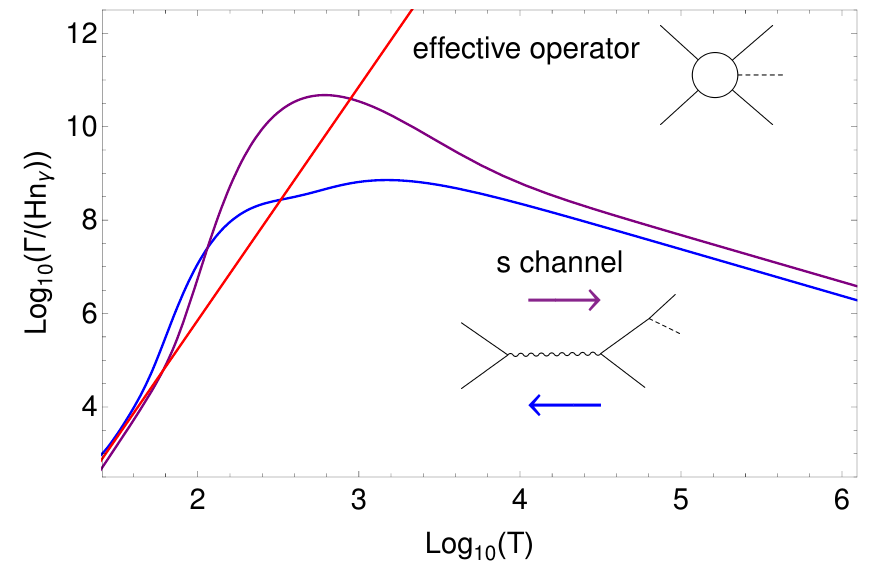}
  \end{minipage}\hfill
  \begin{minipage}[c]{0.32\textwidth}
    \caption{Washout rate calculated in the effective field theory approach and using two UV-complete diagrams. The chosen operator scale is $\Lambda_7= \sqrt{m_F m_B^2}$ with the following heavy mass states: $m_B=1$ TeV and $m_F=2$ TeV.} \label{fig:3}
  \end{minipage}
\end{figure}

Besides the general effective approach, we study also fully UV-completed models causing the effective LNV at low energies to demonstrate the relevancy of the general approach based on effective LNV operators and estimate possible uncertainties. Figure \ref{fig:3} depicts the comparison of washout calculated using the effective operator $\mathcal{O}_7$ and the corresponding UV-completed model. Such an approach allows a more precise calculation of the washout and the interplay between LNV and LFV operators. As seen, even when considering just the s-channel contribution, the washout rate of the completed model is higher than the one calculated for the corresponding effective operator. The observation of heavy resonances mediating LNV at the LHC will also rule out baryogenesis mechanisms in a similar fashion \cite{Deppisch2}.

\Acknowledgements
I am grateful to the authors of the original paper \cite{Deppisch}. I especially thank Wei-Chih Huang for help with preparation of part of the content for the poster.

\end{document}